\PassOptionsToPackage{square,numbers,comma,sort&compress,super}{natbib} 
\documentclass[aps,prl,twocolumn,a4paper,superscriptaddress]{revtex4-2}

\usepackage[paperwidth=215mm,paperheight=300mm,centering,hmargin=1.6cm,vmargin=2cm]{geometry}
\usepackage{graphicx}
\usepackage{amsmath}
\usepackage{placeins}

\usepackage{latexsym,stmaryrd}
\usepackage[dvipsnames]{xcolor}
\usepackage[normalem]{ulem}
\usepackage{float}

\usepackage{siunitx} 
\usepackage{chemformula} 
\usepackage{titlesec}
\usepackage{gensymb}

\titleformat{\section}[hang]{\bfseries}{}{1em}{}
\titlespacing*{\section}{0pt}{2.5ex plus 1ex minus .2ex}{0.2ex plus .2ex}
\titleformat{\subsection}[hang]{\bfseries}{}{1em}{}
\titlespacing*{\subsection}{0pt}{2.5ex plus 1ex minus .2ex}{0.2ex plus .2ex}

\renewcommand{\figurename}{\textbf{Fig.}}

\makeatletter
\renewcommand*{\fnum@figure}{{\normalfont\bfseries \figurename~\thefigure}}
\makeatother

\begin{document}

\title{Design, fabrication, and characterization of electrostatic comb-drive actuators for nanoelectromechanical silicon photonics}

\author{Thor August Schimmell Weis}
\email{taswe@dtu.dk}
\affiliation{Department of Electrical and Photonics Engineering, DTU Electro, Technical University of Denmark, Building 343, DK-2800 Kgs. Lyngby, Denmark}

\author{Babak Vosoughi Lahĳani}
\affiliation{Department of Electrical and Photonics Engineering, DTU Electro, Technical University of Denmark, Building 343, DK-2800 Kgs. Lyngby, Denmark}
\affiliation{NanoPhoton - Center for Nanophotonics, Technical University of Denmark, Ørsteds Plads 345A, DK-2800 Kgs. Lyngby, Denmark.}

\author{Konstantinos Tsoukalas}
\affiliation{Department of Electrical and Photonics Engineering, DTU Electro, Technical University of Denmark, Building 343, DK-2800 Kgs. Lyngby, Denmark}

\author{Marcus Albrechtsen}
\affiliation{Department of Electrical and Photonics Engineering, DTU Electro, Technical University of Denmark, Building 343, DK-2800 Kgs. Lyngby, Denmark}

\author{Søren Stobbe }
\email{ssto@dtu.dk}
\affiliation{Department of Electrical and Photonics Engineering, DTU Electro, Technical University of Denmark, Building 343, DK-2800 Kgs. Lyngby, Denmark}
\affiliation{NanoPhoton - Center for Nanophotonics, Technical University of Denmark, Ørsteds Plads 345A, DK-2800 Kgs. Lyngby, Denmark.}

\homepage{}

\date{\today}
\maketitle

\small
\bfseries\noindent
Nanoelectromechanical systems offer unique functionalities in photonics: The ability to elastically and reversibly deform dielectric beams with subwavelength dimensions enable electrical control of the propagation of light with a power consumption orders of magnitude below that of competing technologies, such as thermo-optic tuning. We present a study of the design, fabrication, and characterization of compact electrostatic comb-drive actuators tailored for integrated nanoelectromechanical silicon photonic circuits. Our design has a footprint of \SI[parse-numbers = false]{1.2\times10^3}{\micro\meter^2} and is found to reach displacements beyond \SI{50}{\nano\meter} at \SI{5}{\volt} with a mechanical resonance above \SI{200}{\kilo\hertz}, or, using different spring constants and skeletonization, a mechanical resonance above \SI{2.5}{\mega\hertz} with displacements beyond \SI{50}{\nano\meter} at \SI{28}{\volt}. This is sufficient to induce very large phase shifts and other optical effects in nanoelectromechanical reconfigurable photonic circuits.

\normalfont

\section{Introduction}
Microelectromechanical systems (MEMS) have become ubiquitous in communication systems \cite{RFMEMSReview} and sensing \cite{MEMSsensorReview}, and large-scale MEMS micromanipulators have been employed in areas ranging from biomedical applications \cite{microInjector} to scanning probe microscopy \cite{cubeStage}. A central element of MEMS is actuation, and several mechanisms can be employed, such as electrostatic, piezoelectric, thermal, magnetic, or electrochemical actuation. One of the most commonly used mechanisms is electrostatic actuation \cite{actuationMechanisms}, where the electrostatic force arising from the capacitive coupling between two electrodes is used to transduce between the electrical and mechanical energy domains. Electrostatic actuators scale well with large surface areas and small gap sizes, are very power efficient, and can be realized with relatively simple nanofabrication involving lithography, etching, and underetching \cite{Hane}. The simplest geometry of an electrostatic actuator is that of two parallel electrodes of which at least one is mechanically compliant, and an applied voltage generates an electrostatic force similar to that of a parallel-plate capacitor. However, the electrostatic force in such direct electrostatic actuators grows nonlinearly with both displacement and voltage, and they are therefore prone to electrostatic pull-in instabilities \cite{Senturia,pull-in}. This has motivated the development of sliding actuators for which the electrodes are displaced laterally and slide next to each other, such that the electrostatic force, in the ideal case, is independent of displacement. This makes sliding actuators resilient against the pull-in instability, which limits the practical travel range of direct electrostatic actuators, in the direction of actuation. The actuation force can then be increased by constructing arrays of sliding electrodes, and such actuators are referred to as comb-drive actuators \cite{Tang, Hane, Fedder, largeDisplacment, cubeStage}.

The miniaturization of MEMS combined with advances in nanofabrication technologies have led to the vision of nanoelectromechanical systems (NEMS), which are not just scaled-down versions of MEMS but offer a wide range of new and different functionalities. For example, NEMS have less mass and higher surface-to-volume ratio compared to their MEMS counterparts, making them ideal for many sensing applications in fields such as biochemistry \cite{bioFET, Li2007}, thermal sensing \cite{thermalTorsion}, and particle mass spectroscopy \cite{massSpectroscopy}.

The prospects of NEMS are particularly strong within photonics because NEMS enable direct integration with silicon photonics or other planar photonics platforms where elastic deformation of photonic waveguides through electrostatic actuators built in the photonic device layer allows for filtering \cite{wuRing}, control of phase \cite{quackPhase}, and routing of guided light \cite{Switch,circuitCrossing}. This evidences great promise for photonic NEMS as a fast, compact, and energy-efficient platform for reconfigurable photonics, which can potentially outperform commonly used approaches such as thermo-optic technology on essentially all figures of merit \cite{phaseshifter, MORPHIC}.

Most previous works on electrostatic NEMS have so far assumed that NEMS are governed by the same physics as MEMS, but it turns out that both the electrostatic and the material properties change at the nanoscale. For example, the electrostatic force in thin actuators is dominated by fringing fields rather than parallel fields \cite{ktso}. It is therefore essential to establish accurate models and design rules for electrostatic NEMS. This in turn requires a direct comparison between experiment and theory for the displacement, operating bandwidth, and several other important parameters, which has so far been missing in the literature.

The ideal NEMS comb drive should have a compact footprint, low driving voltage, large travel range, and high operating frequency. However, the extreme sensitivity of optical fields to the displacement of boundaries of materials with high refractive indices such as silicon, means that the travel range is often less important for photonic NEMS. Balancing these parameters calls for a compact design with a high density of comb fingers.

Here we focus on developing compact comb-drive actuators with sub-micrometer displacements and high operating frequencies fabricated on silicon on insulator (SOI) platforms with a device-layer thickness of \SI{220}{\nano\meter}, i.e., the standard thickness for silicon photonics. To this end, we use a compact folded-cantilever spring design together with a battering-ram body, as illustrated in Fig.~\ref{fig1}(a). The gap between opposing comb fingers, $g$, (as shown in Fig.~\ref{fig1}(b)) determines the density of the comb fingers and strikes a balance between the electrostatic force at a given actuation voltage and the risk of lateral collapse and stiction.

\section{Theory}
\begin{figure*}[htbp]
\centering
 \includegraphics[width=\textwidth]{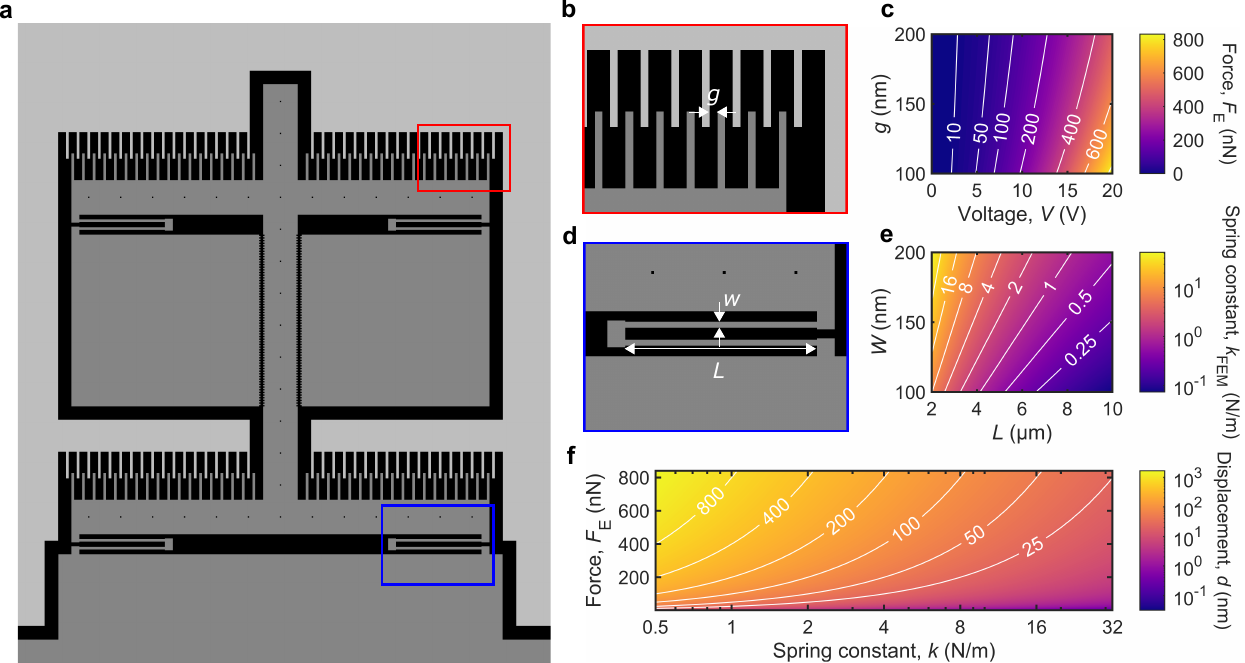}
    \caption{\textbf{Modeling and design of comb-drive actuators for nanoelectromechanical silicon photonics.} (a) Schematic of a battering-ram comb-drive actuator. The two tones of gray indicate areas of different electric potentials during operation. The movable parts are suspended by springs attached to islands large enough to avoid underetching. (b) Zoom-in of comb-drive fingers with gap size $g$. (c) Contour plot of the electrostatic force of a comb drive with 72 finger pairs as a function of gap size and actuation voltage. (d) Zoom-in of a folded-cantilever spring with beam width $W$ and length $L$. (e) Logarithmic contour plot of the spring constant of four folded-cantilever springs in parallel as a function of beam width and length (the complete comb drive includes four springs). (f) Logarithmic contour plot of the comb-drive displacement as a function of the total spring constant of the four-spring suspension and the total electrostatic force.}
    \label{fig1}
\end{figure*}

The electrostatic force, $F_\text{E}$, acting on an actuator is found as the derivative of the stored electrostatic potential energy \cite{Fedder} $F_\text{E} = \frac{1}{2}\frac{\partial C}{\partial x}V^2$. In MEMS, the capacitance between fingers is often well approximated by the parallel-plate approximation due to the large thickness of the device layer compared to the gap between fingers. However, due to the thin device layer of our platform, fringing fields contribute significantly to the capacitance. While this changes the scaling of the capacitance with respect to $g$, the resulting electrostatic force remains largely independent of the displacement of the comb drive \cite{ktso}. Figure~\ref{fig1}(c) shows a contour plot of the electrostatic force generated by a comb drive with 72 finger pairs as a function of voltage, $V$, and gap size, $g$, calculated using finite-element modeling.
Our comb-drive design is suspended by four folded cantilevers of width $W$ and length $L$ as shown in Fig.~\ref{fig1}(d). Each folded cantilever consists of two long beams connected by a short truss. Assuming stiff connecting trusses, these folded cantilevers behave as four parallel connections of two guided cantilevers in series, resulting in a total spring constant estimated by Euler-Bernoulli beam theory \cite{Senturia, beamTheory} as
\begin{equation}\label{kAnalytical}
    k_\text{EB} = 2ET\left(\frac{W}{L}\right)^3,
\end{equation}
where $E = \SI{169}{\giga\pascal}$ is Young's modulus of silicon in the $\langle 110\rangle$ direction \cite{YoungsModulus} and $T = \SI{220}{\nano\meter}$ is the thickness of the device layer. This analytical model provides a simple initial estimate, but it cannot accurately capture the behavior of the system. A numerical finite-element approach is necessary to more precisely estimate the spring constants. Figure~\ref{fig1}(e) shows a contour plot of the spring constant, $k_\text{FEM}$, of four folded-cantilever springs in parallel calculated with the finite-element method as a function of the width, $W$, and the length, $L$, of the folded cantilevers. This relation is important for designing comb-drive actuators because the spring constant directly impacts the trade-off between low-voltage and high-frequency operation.
The displacement, $d$, of a comb drive in steady state is determined by balancing the electrostatic force against the spring force,
\begin{equation}
    d = \xi V^2,
\end{equation}
where we define the comb-drive constant $\xi = \frac{1}{2k}\frac{\partial C}{\partial d}$. Figure~\ref{fig1}(f) shows a contour plot of the steady-state displacement of comb drives with spring constants ranging from $0.5$ to \SI{32}{\newton\per\meter} and a range of forces varying as specified in Fig.~\ref{fig1}(c).

Figure~\ref{fig1} illustrates the design space for comb-drive actuators with thin device layers and illustrates the trade-off between displacement, spring constant, and actuation voltage over a parameter range that is typical for silicon photonics. This serves as a design guide for comb-drive actuators in silicon photonics and enables identifying the geometrical parameters to yield a desired performance. However, this model neglects a number of effects of both theoretical and practical nature. First, the small aspect ratio of silicon photonic NEMS makes them more sensitive to unwanted out-of-plane forces, such as the levitation force or the out-of-plane pull-in instability, depending on whether the substrate is grounded or floating \cite{levitation}. Additionally, a number of works have reported a significant reduction \cite{ESize, EsizeMeasurement} of Young's modulus of silicon relative to the bulk value for suspended devices with critical dimensions below \SI{350}{\nano\meter}. In practice, we focus our work on springs with a width of \SI{200}{\nano\meter} as this ensures that the in-plane spring constant exceeds the out-of-plane spring constant for a single suspended beam while minimizing unwanted effects associated with smaller critical dimensions. In any case, this calls for careful experimental studies, which we present in the following.

\begin{figure*}[htbp]
    \centering
    \includegraphics[width=\textwidth]{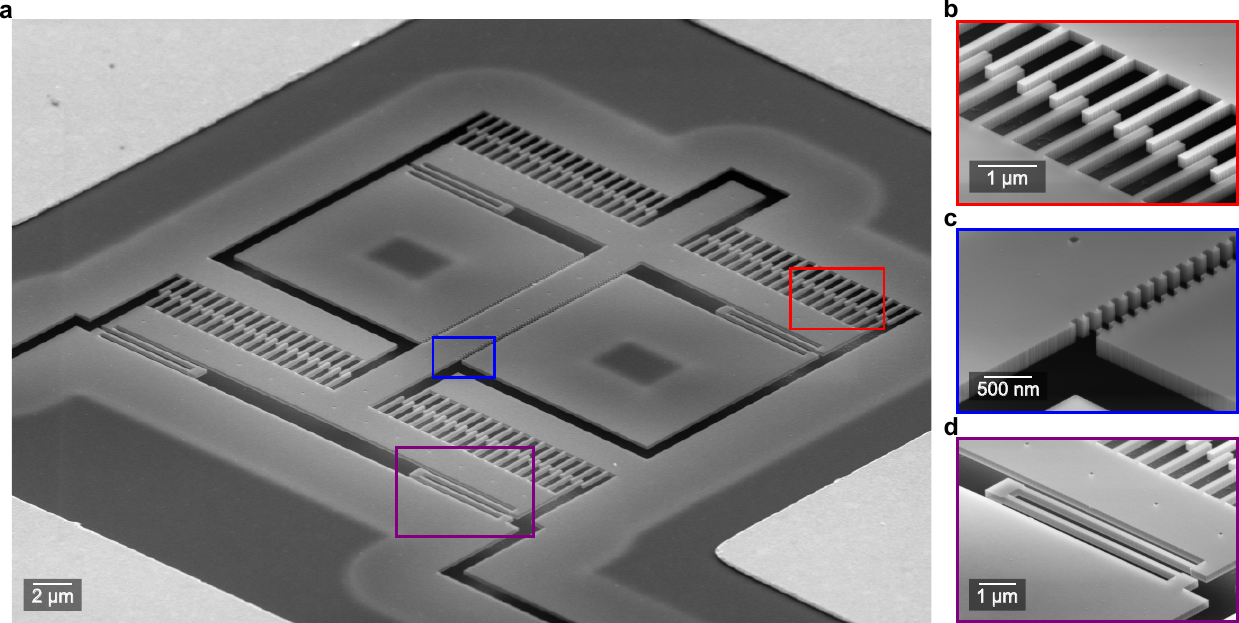}
    \caption{\textbf{Fabricated comb-drive actuator.} (a) Scanning electron micrograph (\SI{60}{\degree} tilted) of a battering-ram comb-drive actuator. Light gray areas of the comb drive are suspended, while dark gray areas rest on the buried oxide layer. White areas in the corners of the micrograph are metallic electrical contacts. (b) Zoom-in of comb drive fingers. (c) Zoom-in of periodic features etched along the spine of the comb drive to enable measuring displacements by SEM with single-nanometer precision \cite{selfAssembly}. (d) Zoom-in of a folded-cantilever spring.}
    \label{fig2}
\end{figure*}

\section{Fabrication}
We fabricate comb drives with $N=72$ finger pairs, a gap between fingers of $g=\SI{200}{\nano\meter}$, a spring width of $W=\SI{200}{\nano\meter}$, and a spring length varying between $L=[2.55$, 3.21, 4.04, 5.09, 6.42, 8.08, 10.18] $\si{\micro\meter}$ corresponding to spring constants $k_\text{EB}=[0.5$, 1, 2, 4, 8, 16, 32] $\si{\newton\per\meter}$ according to (\ref{kAnalytical}). We have also investigated finger gaps down to 50 nm, but we find that this often leads to in-plane collapses, and we therefore only present the results on 200 nm finger gaps.

Our comb drives are fabricated from nominally undoped SOITEC SOI wafers with a \SI{220}{\nano\meter} device layer, a resistivity of $\rho \sim \SI{10}{\ohm\centi\meter}$ specified by the manufacturer, and a \SI{2}{\micro\meter} buried oxide (BOX) layer. The substrate is spin-coated (ALLRESIST CSAR 62), and patterned by electron-beam lithography (JEOL JBX-9500FSZ \SI{100}{\kilo\volt}). The pattern is then transferred into the device layer by reactive-ion etching (SPTS-Pegasus). Afterwards, metal contacts are formed by liftoff using a negative UV-sensitive resist (AZ nLOF 2020) exposed with a maskless aligner (Heidelberg MLA100). Metal contacts are defined by electron-beam evaporation (Ferrotec Temescal FC-2000) of a \SI{5}{\nano\meter} Cr adhesion layer followed by a \SI{200}{\nano\meter} Au film. As the last step, the structures are released by a selective etch of the BOX layer with a temperature- and pressure-controlled vapor phase HF etch (SPTS Primaxx uEtch).

Figure \ref{fig2}(a) shows a scanning electron microscope (SEM) image of a fabricated comb drive. The light gray areas in the corners of the image show the two metal contacts used to drive the actuator. The dark gray areas are the silicon device layer supported by the BOX layer underneath, and the gray areas are the sections of the device layer that have been released from the BOX layer by isotropic selective underetching. Built-in stress in the device layer strains the comb drive after underetching, and while the springs help absorb some of this strain, the built-in stress has caused the comb drive to buckle slightly up along it's spine. This buckling causes the comb-drive fingers to disengage by a few nanometers, which in turn slightly lowers the electrostatic force. While the effect in this case is likely negligible, it could become significant if the spine is extended to accommodate more rows of fingers, which would require additional stress-release structures or other types of stress management. Figure~\ref{fig2}(b) shows a zoom-in of the comb-drive fingers. The difference in brightness between the two sets of fingers is caused by SEM charging effects and shows that they are electrically isolated from each other. Figure~\ref{fig2}(c) shows a periodic scale with a \SI{200}{\nano\meter} period that runs along the spine of the comb drive and along the two anchored islands. This scale can be automatically located in an SEM image by searching for its known periodicity, and it allows measuring the displacement of the comb drive from scanning electron micrographs with single-nanometer precision by Fourier analysis \cite{selfAssembly}. Figure~\ref{fig2}(d) shows a fabricated spring. Besides their mechanical function, the springs at the bottom of the comb drive serve as electrical connections between a nearby metal contact and the moving part of the comb drive.

Conventionally, comb drive actuators are doped, which for many applications can be done without introducing additional complexity to the fabrication process, whether by working with pre-doped SOI wafers or by fabricating actuators in polysilicon, which can be deposited with doping by low-pressure chemical vapor deposition (LPCVD) \cite{PSOI}. However, in photonic NEMS, indiscriminate doping of the device layer induces optical losses, and as such, a patterned doping process is needed, which increases the complexity of the fabrication process \cite{MORPHIC}. This motivates an investigation of when it is necessary to dope comb-drive actuators. One concern with undoped silicon device layers is the formation of Schottky barriers where metal contacts interface with the intrinsic silicon, as these barriers behave as diodes \cite{Sze}. A minimum of two electrical contacts are needed to operate a comb drive, implying that the equivalent electrical circuit has two opposing Schottky diodes in series, such that one Schottky diode is always reversely biased.
This is normally not an issue for electrostatic actuators during quasi-static operations since the reverse leakage current is usually enough to drive the actuator, but the high impedance of the reversely biased diode, and thus the increased circuit response time (RC-time) can restrict the dynamic response of the actuator to frequencies far below the mechanical resonance frequencies \cite{MORPHIC}.

\begin{figure*}[htbp]
    \centering
    \includegraphics[width=\textwidth]{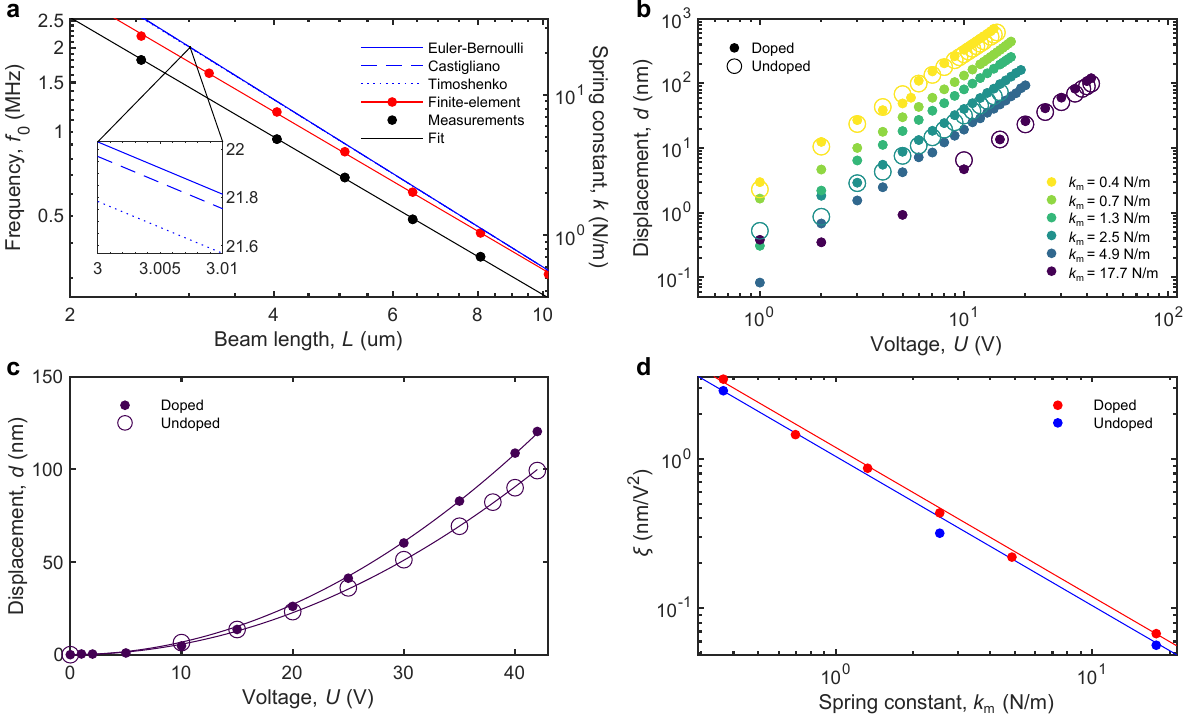}
    \caption{\textbf{Measured dynamic and steady-state response of comb drive actuators.} (a) Measured resonance frequencies and extracted spring constants of doped comb drives as a function of the length of the cantilever beams. The solid, dashed, and dotted blue lines represent analytical models of the spring constant using Euler-Bernoulli beam theory, Castigliano's theorem, and Timoshenko beam theory, respectively. The solid red line shows a power law fitted to the finite-element results. That power law is subsequently fitted to the experimental data by keeping the exponent fixed and the prefactor free. The resulting fit is shown as the solid black line and shows excellent agreement with the experiments. (b) Actuator displacement as a function of the applied voltage on log-log axes for doped and undoped comb drives with suspensions of different measured spring constants, $k_\text{m}$, showing slightly more displacement for the doped comb drives compared to their undoped counterparts. (c) Displacement as a function of actuation voltage on linear axes for doped and undoped comb drives with a measured spring constant of $k_\text{m}= \SI{17.7}{\newton\per\meter}$. (d) Comb-drive constants of nominally identical comb drives, with and without doping, as a function of measured spring constants, $k_\text{m}$.}
    \label{fig3}
\end{figure*}

\begin{figure*}[htbp]
    \centering
    \includegraphics[width=\textwidth]{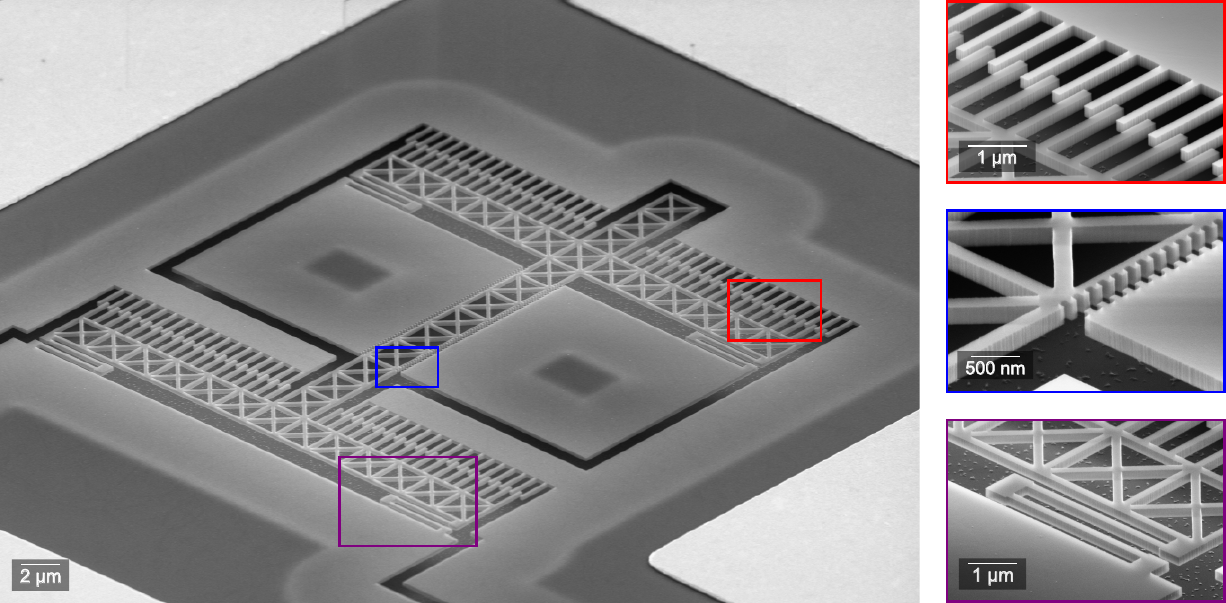}
    \caption{\textbf{Skeletonized comb drive for high-speed actuation.} Scanning electron micrograph of skeletonized battering-ram comb-drive actuator. The skeletonization reduces the movable mass, thus resulting in an actuator with a higher resonance frequency than an equivalent actuator without skeletonization.}
    \label{fig4}
\end{figure*}

\section{Characterization of dynamic performance}
To study the effect of doping, we fabricate two sets of nominally identical comb drive actuators and dope one set by diffusion-doping using boron-doped silica to obtain a resistivity of $\rho = \SI[parse-numbers = false]{6\times10^{-4}}{\ohm\centi\meter}$.
The doped comb drives are not limited by the electrical impedance, and we can excite their mechanical resonances by applying an alternating-current (AC) signal. We measure the dynamic response by applying the AC drive in-situ in an SEM. To illustrate the capabilities of this in-situ technique, we refer to supplementary \textbf{Visualization 1} for an animation composed of single SEM images of the comb-drive obtained for different values of an applied constant bias. Returning to the AC-measurements, we measure the amplitude of the blur \cite{Hane} of the periodically spaced void features in the body of the comb drives. 
Applying a voltage across two electrodes on a sample in an SEM deflects the electron beam much like the deflector coils do. With a direct-current (DC) drive, this can be corrected for by adjusting the deflector coils, but with AC signals with frequencies above the scan rate of the SEM, the beam deflection shows up as image blur that makes it harder to measure the blur created by the mechanical vibrations of the comb drive. We find that keeping the AC amplitude in the 0.1-\SI{1}{\volt} range reduces the beam deflection to an acceptable level, but such a voltage range is too low to noticeably actuate the comb drives. We therefore apply a DC bias to operate the comb drive at a working point with larger $\frac{\mathrm{d} d}{\mathrm{d} V}$ in order to increase the sensitivity. Since the electrostatic force acting on the comb drives scales with the square of the voltage, $F_\text{E}\propto V^2$, the force generated by an AC signal with a DC bias, $V = V_0 + A\sin{(\omega t)}$, has two frequency components
\begin{equation}
    F_\text{E}\propto V^2 = V_0^2 + 2V_0A\sin{(\omega t)} + A^2\frac{1-\cos{(2\omega t)}}{2},
\end{equation}
where $V_0$ is the DC bias, $A$ is the AC amplitude, $\omega$ is the angular frequency, and $t$ is time. For an AC signal with zero DC bias, the force will oscillate at twice the frequency of the input signal, but for $V_0>>A$ the frequency of the generated force is dominated by a single frequency component, $\omega$.
The increased sensitivity enabled by the DC bias allows us to image the oscillating comb drives without significant interference from beam deflections. The mechanical resonance frequencies corresponding to the observed peak amplitudes of the image blur are plotted in Fig.~\ref{fig3}(a) as a function of the beam length, $L$. From the resonance frequency, we extract the measured spring constant as $k_\text{m} = \left(2\pi f_0\right)^2m$, where $f_0$ is the resonance frequency and $m$ is the mass of the comb drive. The plot also shows the prediction from three analytical solutions commonly employed in the design and analysis of MEMS in order to probe their validity in the regime of photonic NEMS. The first model is Euler-Bernoulli beam theory, cf.\ Eq.~(\ref{kAnalytical}), but we include also calculations using Castigliano's theorem \cite{Fedder}, which takes the connecting trusses into account,
\begin{equation}
    k_\text{C} = 2ET\left(\frac{W}{L}\right)^3\frac{4L+\frac{W^3}{W^3_\text{T}}L_\text{T}}{4\left(L+\frac{W^3}{W^3_\text{T}}L_\text{T}\right)}.
\end{equation}
Finally, we consider Timoshenko beam theory \cite{beamTheory} which takes shearing of the spring beams into account,
\begin{equation}
    k_\text{T} = \left(\frac{L}{2\kappa TWG}+\frac{L^3}{2ETW^3}\right)^{^-1},
\end{equation}
where $\kappa=\frac{5}{6}$ is the Timoshenko shear coefficient for rectangular cross section and $G=\SI{79}{\giga\pascal}$ is the shear modulus of silicon in the $\langle 110\rangle$ direction \cite{YoungsModulus}. All three models give very similar results in reasonable agreement with the measured eigenfrequencies although with significant deviations in both the scaling and the prefactor. To ensure a comprehensive analysis, we also include finite-element calculations and compare them with the experimental and analytical models presented in Fig.~\ref{fig3}(a). The spring constants calculated with the finite-element method capture the observed scaling but with a different prefactor, which could be due to built-in stress, or it may originate from the fabricated springs having a lower Young modulus than expected due to surface effects associated with the small critical dimensions in our devices \cite{ESize, EsizeMeasurement}. To find the exact reduction in stiffness, the numerically calculated spring constants are fitted with a power law as indicated with a red curve in Fig.~\ref{fig3}(a). The fitted exponent of this power law is then kept fixed as the power law is then fitted to the experimental data with the prefactor as a free parameter. This results in an excellent agreement between theory (black line) and experiment (black dots). The prefactor extracted by this method corresponds to a $34\pm1\%$ drop in stiffness compared to the expected value.

\section{Characterization of steady-state behaviour}
We measure the displacement of the actuators as a function of applied DC voltage using image analysis of SEM images aquired with zero tilt angle of the periodic structures shown in Fig.~\ref{fig2}(c). The result is shown in Fig.~\ref{fig3}(b). For linear springs, the displacement scales with the square of the voltage, and all displacement curves would be parallel lines on log-log axes, which is confirmed by our experiments. While subtle, the displacement at a given applied voltage is slightly less for the undoped comb drives as compared to the doped devices. This is more apparent in a plot with linear axes as shown in Fig.~\ref{fig3}(c). The doped comb drives displace 20\%, 37\%, and 20\% more than undoped nominally identical comb drives for measured spring constants of $k_\text{m}$ = \SI{0.4}{\newton\per\meter}, \SI{2.5}{\newton\per\meter}, and \SI{17.7}{\newton\per\meter}, respectively. This extra displacement could be due to a lower Young modulus of the highly doped springs. However, it has been reported that even heavy doping only reduces Young's modulus of silicon by up to 3\% \cite{YoungsModulus}. Another explanation could be a reduced Debye screening in the undoped silicon \cite{DebyeBeliever}, which would result in a smaller differential capacitance, $\frac{\partial C}{\partial d}$. To the best of our knowledge, this effect has not yet been experimentally investigated in electrostatic actuators. In any case, this shows that (nominally) undoped silicon works very well for DC operation of photonic NEMS.

We now turn to the extraction of the comb-drive constant, $\xi = \frac{1}{2k}\frac{\partial C}{\partial d}$, which we have defined as the proportionality between $V^2$ and $d$, i.e., it can readily be extracted by fitting a quadratic function to the measured displacement curves in Fig.~\ref{fig3}(b). The extracted comb-drive constants are plotted as a function of measured spring constants, $k_\text{m}$, in Fig.~\ref{fig3}(d), from which we can directly determine the differential capacitances. We find $\frac{\partial C}{\partial d} = \SI[separate-uncertainty=true,multi-part-units=single]{2.4(0.1)}{\nano\farad\per\meter}$ for the doped comb drives, and $\frac{\partial C}{\partial d} = \SI[separate-uncertainty=true,multi-part-units=single]{2.1(0.1)}{\nano\farad\per\meter}$ for the undoped comb drives, where we have assumed identical spring constants for doped and undoped actuators.

\section{High-frequency actuators}
For high-speed applications, a comb drive with a high mechanical resonance frequency is required and doping is required to minimize the circuit impedance. The resonance frequency of a comb drive can be increased either by using stiffer springs or by reducing the mass of the comb drive. Skeletonizing a comb drive reduces mass at the cost of decreasing the stiffness of the comb-drive body. Figure~\ref{fig4} presents a skeletonized comb drive based on our proposed design. For this comb drive, we measure a resonance frequency of \SI{2.7}{\mega\hertz} when suspended by a spring constant of \SI{17.7}{\newton\per\meter}, which, to the best of our knowledge, is the highest resonance frequency ever reported for a comb-drive actuator.

\section{Conclusion}
In conclusion, we propose and experimentally validate a comb-drive actuator designed for reconfigurable photonic integrated circuits. Our design concept allows for a wide range of customization balancing the desired performance and trade-offs between speed, displacement, and actuation voltage. For example, by choosing a low spring constant of $k_\text{m}=\SI{0.4}{\newton\per\meter}$, a displacement of \SI{50}{\nano\meter} can be reached with actuation voltages below \SI{5}{\volt}. Such a displacement is sufficient to reconfigure tunable photonic components such as phase shifters \cite{phaseshifter} and directional couplers \cite{MORPHIC} with low actuation voltages while still operating at frequencies above \SI{200}{\kilo\hertz}. Moreover, the proposed actuator can actuate stiff springs, enabling operation at MHz frequencies with moderate actuation voltages thanks to the tightly arranged interdigitated comb fingers. It is important to note that in a practical application, the displacement and the operating frequency of the comb-drive actuator can be reduced depending on the mass and stiffness of an attached external load. We observe that doping is necessary for high-speed applications, as the high impedance of Schottky barriers otherwise results in an electrical response time significantly slower than the mechanical response time.

\section{Acknowledgements}
The authors gratefully acknowledge financial support from
Innovation Fund Denmark (Grant No. 0175-00022 -- NEXUS and Grant No. 2054-00008 -- SCALE),
the Danish National Research Foundation (Grant No. DNRF147 -- NanoPhoton),
Independent Research Fund Denmark (Grant No. 0135-00315 -- VAFL),
the European Research Council (Grant No. 101045396 -- SPOTLIGHT),
and the European Union's Horizon research and innovation programme (Grant No. 101098961 -- NEUROPIC).

\section{Data availability}
Data is available upon reasonable request.

\section{Competing financial interests}
The authors declare no competing financial interests.

\scriptsize

\bibliographystyle{naturemag-etalnoitalics.bst}
\bibliography{Bibliography}

\end{document}